\documentclass[12pt]{article}
\usepackage{epsf}
\usepackage{amsfonts}
\title{An $f(R)$ model for dark matter: rotation curves and gravitational lensing}
\author{F. Shojai and A. Shojai\\
Department of Physics, University of Tehran, Tehran, Iran.} 
\date{}
\usepackage{graphicx}
\begin{document}
\maketitle
\begin{abstract}
There should be two ways to describe the flat rotation curves of galaxies and cluster of galaxies. Either one can introduce a dark matter component for the matter filling the halo, or by modifying the gravity theory and give the flat rotation curve a geometrical nature. Here we adopt an $f(R)$ model suitable for describing the effect. After matching the solution with the exterior solution, the effective density, radial and tangential pressures are obtained. Then the energy conditions and lensing effect is investigated.
\end{abstract}
\section{Introduction}
The concept of dark matter is introduced first by Zwicky in 1933 \cite{z} as the missing mass problem. He observed that the moving galaxies in Coma cluster move faster than what is  predicted by considering the visible mass of the cluster. Using Virial theorem to estimate the dynamical mass of Coma cluster,
he determined that there is some discrepancy in the observed and computed masses. The next related observation   
is  the rotation curves of spiral galaxies\cite{rubin}. the rotation curves can be determined by measuring the rotational velocities of stars in the optical region of galaxies. The same holds for clusters of galaxies. For non--visible parts of galaxies the frequency shift of $21cm$ line of neutral hydrogen clouds is used to determine the velocity. 

These observations show that the rotation curve remains nearly flat at enough large distance from the center. Both of these observations usually are explained by assuming a spherical halo of \textit{dark matter} around the astronomical objects. Moreover in describing the stability of disk galaxies\cite{os} and also in the structure formation at the cosmological scales\cite{pe}, dark matter component plays an essential role.

The concept of dark matter component is basically a  prediction of general theory of relativity applied to the astronomical systems. In fact there are two roads for cosmologists and astronomers for describing the observed facts, first, adding an exotic component to the matter content of the universe\cite{bru}, and second, modifying Einstein's theory of gravity in such a way that the correct results can be obtained using the observed ordinary matter\cite{tev,sot}.

In the first viewpoint, the cold dark matter (CDM) is successful to describe CMB anisotropies and the large scale structure of the universe. But it suffers from some problems such as Cusp--Core problem\cite{mikh}. By considering the dark matter component as a perfect fluid,  it is straightforward to show that applying the weak energy condition (WEC) and matching with the Schwarzschild metric at the halo surface, make the dark matter pressure-less\cite{bha}. Therefore the authors in \cite{bha} assumed that the dark matter fluid is not ideal and that it's pressure has different radial and tangential components. By suggesting a variety of equations of state for dark matter and using the linear field approximation of general relativity, they produce the flat part of the rotation curve. 

This can also be achieved using a scalar field\cite{mat} as dark matter. An exponential potential for the scalar field is not only able to produce  a constant orbital velocity but also can  make  the energy conditions to be satisfied\cite{mat}. Furthermore by introducing a novel scalar field the authors of \cite{su} show that a numerical analysis of Einstein's equations gives the correct rotation curve at arbitrary radius. This also solves the Cusp-Core problem.
It has to be noted that taking the flat rotation curve as input, the authors of \cite {rahman} find the space-time metric and the equation of state of dark matter as a prefect fluid. But the solution is not matched to the exterior   Schwarzschild metric.

In the latter viewpoint, there isn't any dark matter component and  the observations are explained by  geometrical effects. This means that to describe flat rotation curves, one uses an extended theory of gravity without introducing dark matter. 
For example, $f(R)$ theories\cite{sot} which are the simplest modifications of general relativity can predict correct rotation curves. Choosing different forms for $f(R)$, one can investigate the dark matter problem \cite{dmf,sob,as}. The authors of \cite{sob} have found the appropriate form of $f(R)$ admitting a solution in the form of a perturbed Schwarzschild metric for the dark matter halo. This leads to a rotation curve which approaches to a constant at  large enough radii from the center of galaxy. Also in \cite{as}, it is shown that a constant orbital velocity leads to $f(R)\sim R^n$ in which $n$ depends to the velocity.

In the above mentioned references \cite{dmf,sob,as} additional assumptions are made to obtain the form of $f(R)$, but here  we propose an $f(R)$ modification of Einstein's gravity suitable for the flat region of rotation curves of galaxies without any additional assumptions.
To do this, in section 2 we first briefly review the field equations of $f(R)$ theory in the form of Einstein's equations with an effective energy--momentum tensor. Then we shall use an exact static spherically symmetric solution of $f(R)$ theory obtained in \cite{khod}, and see that it is not possible to match it to the exterior  
Schwarzschild metric at the surface of the halo. Due to small circular velocities of objects in the halo, we follow an approximate method  to find another solution of $f(R)$ theory suitable for describing rotation curves. The boundary conditions at the surface of halo are used to fix the arbitrary constants of the solution. In this way the effective density and pressure of dark matter are obtained.

Finally in section 3 we discuss the gravitational lensing properties of this solution.
\section{Rotation curves and $f(R)$ gravity}
To describe the motion of a test particle in the halo of a spiral galaxy, it is usually assumed that the space--time metric is spherically symmetric and static, i.e.:
\begin{equation}
ds^2=A(r)dt^2-B(r)dr^2-r^2d\Omega
\label{met}
\end{equation}
The geodesic equation for such a particle rotating around the center of the galaxy in a circular orbit leads to:
\begin{equation}
v_{\textrm{\tiny tg}}^2=\frac{r^2}{A}\frac{d\phi^2}{dt^2}=\frac{r}{2A} \frac{dA}{dr}
\label{v}
\end{equation}
in which $v_{\textrm{\tiny tg}}$ is the tangential velocity of the test particle and we are using the units in which $c=1$. It is a well known fact that the rotation curves of the spiral galaxies tends to a constant tangential velocity in the interval
\begin{equation}
2.3\times10^{-4}<v_{\textrm{\tiny tg}}<8.3\times10^{-4}
\end{equation}
Using relation (\ref{v}), one could get:
\begin{equation}
A(r)\sim A_0 \left(\frac{r}{r_{\textrm{\tiny opt}}}\right)^\alpha
\label{g00}
\end{equation}
in the flat region of the rotation curve, where $A_0$ is an arbitrary constant, $\alpha=2v^2_{\textrm{\tiny tg}}$ is a small dimensionless parameter and
\begin{equation}
1.08\times10^{-7}<\alpha<1.38\times10^{-6}
\end{equation}
$r_{\textrm{\tiny opt}}$ represents the optical or luminous radius of the galaxy. This form of the time--time component of the metric is different from the Schwarzschild one and thus it is not correct to assume that all the matter in the galaxy is the luminous one at its center. This is usually supposed to be an argument for believing in dark matter filling the halo of spiral galaxies.  

An alternative way is to describe the rotation curves as a geometrical effect, using extended gravity models such as $f(R)$ theories. The action functional of $f(R)$ theories is usually written as:
\begin{equation}
{\cal A}=\int d^4x\sqrt{-g}(f(R)+\kappa{\cal L}_m)
\end{equation}
leading to the following field equations:
\begin{equation}
F(R)R_{\mu\nu}-\frac{1}{2}f(R)g_{\mu\nu}-\nabla_\mu\nabla_\nu F(R)+g_{\mu\nu}\nabla^\alpha\nabla_\alpha F(R)=-\kappa T_{\mu\nu}
\label{freq}
\end{equation}
in which $F(R)=df/dR$ and $T_{\mu \nu}$ is the energy--momentum tensor of matter. 

Using the line element (\ref{met}), the vacuum field equations would result in \cite{mul}:
\begin{equation}
2\frac{F}{X}\frac{dX}{dr}+\frac{r}{X}\frac{dF}{dr}\frac{dX}{dr}-2r\frac{d^2F}{dr^2}=0
\label{e1}
\end{equation}
and
\begin{equation}
\frac{d^2A}{dr^2}+\left(\frac{1}{F}\frac{dF}{dr}-\frac{1}{2X}\frac{dX}{dr}\right)\left(\frac{dA}{dr}-\frac{2A}{r}\right) -\frac{2A}{r^2}+\frac{2X}{r^2}=0
\label{e2}
\end{equation}
where $X(r)=A(r)B(r)$. The Ricci scalar is also given by:
\begin{equation}
R=-\frac{2}{r^2}+\frac{1}{X}\left [ \frac{d^2A}{dr^2}+\frac{4}{r}\frac{dA}{dr} -\frac{1}{X}\frac{dX}{dr}\left ( \frac{1}{2}\frac{dA}{dr}+\frac{2A}{r}\right )+\frac{2A}{r^2}\right ]
\label{e3}
\end{equation}
 It is interesting to see which kind of effective energy--momentum tensor is given by this model. To do so, we rewrite equation (\ref{freq}) in the form of Einstein's equations \cite{sot}:
\begin{equation}
R_{\mu\nu}-\frac{1}{2}Rg_{\mu\nu}=-\kappa^{\textrm{\tiny eff}}\left (
T_{\mu\nu}+T^{\textrm{\tiny eff}}_{\mu\nu}\right)
\end{equation}
 where $\kappa^{\textrm{\tiny eff}}=\kappa/F$ is the effective gravitational coupling constant and 
 \begin{equation}
 -\kappa T^{\textrm{\tiny eff}}_{\mu\nu}=
\frac{1}{2}(f-RF)g_{\mu\nu}+\nabla_\mu\nabla_\nu F-g_{\mu\nu}\nabla^\alpha\nabla_\alpha F
\end{equation}
Putting this in the form of a perfect fluid energy--momentum tensor and inserting the metric (\ref{met}), one can show that:
\begin{equation}
 -\kappa \rho^{\textrm{\tiny eff}}=
\frac{1}{2}(f-RF)+\frac{1}{r_{\textrm{\tiny opt}}^2}\left ( \frac{AF\rq{}\rq{}}{X}-\frac{F\rq{}AX\rq{}}{2X^2} +\frac{F\rq{}A\rq{}}{2X} + \frac{2F\rq{}A}{uX}\right )
\end{equation}
\begin{equation}
-\kappa p_r^{\textrm{\tiny eff}}=
\frac{1}{2}(f-RF)+\frac{1}{r_{\textrm{\tiny opt}}^2}\left (\frac{A\rq{}F\rq{}}{2X}+\frac{2F\rq{}A}{uX}\right )
\end{equation}
\begin{equation}
-\kappa p_t^{\textrm{\tiny eff}}=
\frac{1}{2}(f-RF)+\frac{1}{r_{\textrm{\tiny opt}}^2}\left (\frac{AF\rq{}\rq{}}{X}-\frac{F\rq{}AX\rq{}}{2X^2} +\frac{F\rq{}A\rq{}}{X}+\frac{F\rq{}A}{uX}\right )
\end{equation}
where $ \rho^{\textrm{\tiny eff}}$, $p_r^{\textrm{\tiny eff}}$ and $p_t^{\textrm{\tiny eff}}$ are the effective density, radial and tangential pressures respectively. Also a prime over any quantity represents differentiation with respect to the normalized radial coordinate 
\begin{equation}
u=\frac{r}{r_{\textrm{\tiny opt}}}
 \end{equation}
\subsection{An exact $f(R)$ spherically symmetric solution suitable for dark matter}
Using the exact solutions of \cite{khod} (see the solution I of table 2), a simple solution of vacuum spherically symmetric static character suitable for flat rotation curves can be found. It is:
\begin{equation}
A=A_0u^\alpha
 \end{equation}
\begin{equation}
X=X_0u^\alpha
 \end{equation}
 \begin{equation}
F=F_0u^\ell
 \end{equation}
 where
 \begin{equation}
\ell^2-\left( 1+\frac{\alpha}{2}\right)\ell-\alpha=0
 \end{equation}
 and 
  \begin{equation}
X_0=A_0\left [1-\frac{\alpha(\alpha-1)}{2} +\left(\ell-\frac{\alpha}{2}\right) \left(1-\frac{\alpha}{2}\right) \right]
 \end{equation}
 This leads to the following form of $f(R)$:
\begin{equation}
f(R)=f_0+\frac{F_0{r_{\textrm{\tiny opt}}}^{-l}}{\beta(1-\ell/2)}(\beta R)^{1-\ell/2}
\label{22}
\end{equation}
where $\beta^{-1}=\left(\alpha(\alpha+1)+\ell(\alpha-2)\right)/{(1-\alpha/2)(1+\alpha/2+\ell)}$
and $X_0$, $F_0$ and $f_0$ are arbitrary constants.  With $\alpha\sim 10^{-6}$ one gets $(1-\ell/2)\simeq(1+5\times 10^{-7})=1+v^2_{\textrm{\tiny tg}}$ or $(1-\ell/2)\simeq(0.5-7.5\times10^{-7})$.
 The first choice for $\ell$ essentially leads to the solution given in \cite{as}, while the second is $f(R)\sim\sqrt{|R|}$.
 
 In order to have a meaningful solution, we have to match the solution to the Schwarzschild solution outside the halo, defined as:
 \begin{equation}
 A_{\textit{\tiny out}}=E\left(1-\frac{2\mu}{u}\right)
 \label{aout}   
 \end{equation}
  \begin{equation}
 X_{\textit{\tiny out}}=E
 \label{xout}
 \end{equation}
 where $E$ is a constant, $\mu=GM_{\textit{\tiny total}}/r_{\textrm{\tiny opt}}$ and in terms of the tangential velocity or $\alpha$  can be written as $\mu=\alpha u_s/2$.
  The surface of the halo can be defined as:
 \begin{equation}
 p_r^{\textrm{\tiny eff}}|_{u_s}=\rho^{\textrm{\tiny eff}}|_{u_s}=0
 \label{sur}
 \end{equation}
 where $u_s$ is the normalized radius of the halo.
  
 It can be simply shown that it is impossible to do match the above exact internal solution to the exterior Schwarzschild one.
\subsection{An approximate $f(R)$ spherically symmetric solution suitable for dark matter}
Since the parameter $\alpha$ is small, one can adopt a perturbative solution. Let\rq{}s write:
\begin{equation}
A= A_0(1+\alpha\ln u)
\label{ain}
 \end{equation}
For not so large values of $u$ this is essentially $A_0u^\alpha$. 
  
Introducing 
\begin{equation}
X=e^S  
\end{equation}
 and
 \begin{equation}
 F=e^T 
  \end{equation}
  the equations (\ref{e1}) and (\ref{e2}) are then:
\begin{equation}
2S\rq{}+uS\rq{}T\rq{}-2u(T\rq{}\rq{}+T\rq{}^2)=0  
\end{equation}
  \begin{equation}
  D_1\left(T\rq{}-\frac{1}{2}S\rq{}\right)+D_2+\frac{2}{u^2}e^S=0
  \end{equation}
  where
  \begin{equation}
  D_1=A\rq{}-2\frac{A}{u}=-2\frac{A_0}{u}+\alpha\frac{A_0}{u}(1-2\ln u)+\cdots
  \end{equation}
and
  \begin{equation}
  D_2=A\rq{}\rq{}-2\frac{A}{u^2}=-2\frac{A_0}{u^2}-\alpha\frac{A_0}{u^2}(1+2\ln u)+\cdots
  \end{equation}
  Expanding $S$ and $T$ in terms of powers of $\alpha$:
  \begin{equation}
  S=S_0+\alpha S_1+\cdots;\ \ \ \ \ \ \ T=T_0+\alpha T_1+\cdots
  \end{equation}
  the above equations can be solved order by order. 
  
  In zeroth order the equations are:
  \begin{equation}
2S_0\rq{}+uS_0\rq{}T_0\rq{}-2u(T_0\rq{}\rq{}+T_0\rq{}^2)=0 
\end{equation}
\begin{equation}
-2\frac{A_0}{u}\left(T_0\rq{}-\frac{1}{2}S_0\rq{}\right)-2\frac{A_0}{u^2}+\frac{2}{u^2}e^{S_0}=0
\end{equation}
There are two solutions, first
\begin{equation}
S_0=\textrm{constant}=\ln A_0;\ \ \ \ \ \ T_0=\textrm{constant}
\end{equation}
and second
\begin{equation}
S_0=\textrm{constant}=\ln 2A_0;\ \ \ \ \ \ T_0=\textrm{constant}+\ln u
\end{equation}
The first solution has the property of leading to $F=\textrm{const.}(1+\alpha (\textrm{corrections})+\cdots)$ and thus to $f(R)\sim R+\alpha (\textrm{corrections})+\cdots$. Since this latter solution is Einstein\rq{}s gravity corrected by small terms,  we use this solution in the rest of this paper.

In the first order we have:
\begin{equation}
S_1\rq{}-uT_1\rq{}\rq{}=0
\end{equation}
\begin{equation}
\frac{u}{2}S_1\rq{}-uT_1\rq{}+S_1-\frac{1+2\ln u}{2}=0
\end{equation}
with the solution
\begin{equation}
S_1=\ln u+C_1u^2-\frac{2C_2}{u}-1
\end{equation}
\begin{equation}
T_1=-\ln u+C_1u^2+\frac{C_2}{u}+C_3
\end{equation}
where $C_1$, $C_2$ and $C_3$ are integration constants.

As a result up to first order, one gets:
\begin{equation}
X=A_0+\alpha A_0\left ( \ln u +C_1u^2-\frac{2C_2}{u}-1\right)
\label{x}
\end{equation}
and
\begin{equation}
F=F_0+\alpha F_0\left ( -\ln u+C_1u^2+\frac{C_2}{u}+C_3\right )
\end{equation}
where $F_0=e^{T_0}$.

In order to find the form of $f(R)$ one can use equation (\ref{e3}) to obtain $R$ as a function of $u$ and then obtain  $f(R)$ via integration of $F$.  We have:
 \begin{equation}
\frac{R r_{\textrm{\tiny opt}}^2}{3\alpha}=\frac{1}{u^2}-2C_1
\label{44}
\end{equation}
Integrating $F$ we get:
\[
f(R)=f_0+\int dR F=f_0+\int du \frac{dR}{du}F=
\]
\[
f_0 + \frac{3\alpha F_0}{r_{\textrm{\tiny opt}}^2}\left (1-\frac{\alpha}{2}+\alpha C_3\right)\left ( 2C_1+\frac{R r_{\textrm{\tiny opt}}^2}{3\alpha}\right ) +
\]
\[
\frac{2\alpha^2F_0C_2}{r_{\textrm{\tiny opt}}^2}\left ( 2C_1+\frac{R r_{\textrm{\tiny opt}}^2}{3\alpha}\right )^{3/2}
+\frac{3\alpha^2F_0C_1}{r_{\textrm{\tiny opt}}^2}\ln\left ( 2C_1+\frac{R r_{\textrm{\tiny opt}}^2}{3\alpha}\right )+
\]
\begin{equation}
\frac{3}{2}\frac{\alpha^2F_0}{r_{\textrm{\tiny opt}}^2}\left ( 2C_1+\frac{R r_{\textrm{\tiny opt}}^2}{3\alpha}\right )\ln \left ( 2C_1+\frac{R r_{\textrm{\tiny opt}}^2}{3\alpha}\right )
\label{f}
\end{equation}
 
The above $f(R)$ theory can describe the rotation curves without any dark matter.
In terms of the radial distance, from equations (\ref{x})--(\ref{f}) we get:
\begin{equation}
-\kappa\rho^{\textrm{\tiny eff}}=\frac{f_0}{2}+\frac{\alpha F_0}{r_{\textrm{\tiny opt}}^2}\left( 9C_1-\frac{1}{u^2}\right)
\end{equation}
\begin{equation}
-\kappa p_r^{\textrm{\tiny eff}}=\frac{f_0}{2}+\frac{\alpha F_0}{r_{\textrm{\tiny opt}}^2}\left( 7C_1-\frac{2}{u^2}-\frac{2C_2}{u^3}\right)
\end{equation} 
\begin{equation}
-\kappa p_t^{\textrm{\tiny eff}}=\frac{f_0}{2}+\frac{\alpha F_0}{r_{\textrm{\tiny opt}}^2}\left( 7C_1+\frac{C_2}{u^3}\right)
\end{equation}  
   
We can get some relations between the constants by matching the solution to the exterior Schwarzschild metric at $u_s$. Using (\ref{aout}), (\ref{xout}), (\ref{ain}), (\ref{x}) and satisfying condition (\ref{sur}), we get:
\begin{equation}
\begin{tabular}{l}
$f_0=\frac{\alpha F_0\zeta}{r_{\textrm{\tiny opt}}^2u_s^2}$\\
$C_1=\frac{1}{9u_s^2}-\frac{\zeta}{18u_s^2}$\\
$C_2=\frac{u_s}{18}(\zeta-11)$\\
$E=A_0\left(1+\alpha\left(\ln u_s-\frac{\zeta}{6}+\frac{1}{3}\right)\right)$\\
$\mu=\frac{\alpha u_s}{6}\left ( 1-\frac{\zeta}{2}\right)$\\
\end{tabular} 
\label{const}
\end{equation}
where $\zeta$ is some constant which can be obtained using the Newtonian limit relation $\mu=\alpha u_s/2$. This leads to $\zeta=-4$.

Using the above relations for the constants, the components of energy--momentum tensor have the following form:
\begin{equation}
-\frac{\kappa\rho^{\textrm{\tiny eff}}}{f_0}=\frac{1}{4}\left(\frac{u_s}{u}\right)^2-\frac{1}{4}
\label{rhoeff}
\end{equation}
\begin{equation}
-\frac{\kappa p_r^{\textrm{\tiny eff}}}{f_0}=-\frac{1}{12}+\frac{1}{2}\left(\frac{u_s}{u}\right)^2-\frac{5}{12}\left(\frac{u_s}{u}\right)^3
\end{equation}
\begin{equation}
-\frac{\kappa p_t^{\textrm{\tiny eff}}}{f_0}=\frac{5}{24}\left(\frac{u_s}{u}\right)^3-\frac{1}{12}
\label{pteff}
\end{equation}

\begin{figure}[h]
\centering
\includegraphics[scale=0.5]{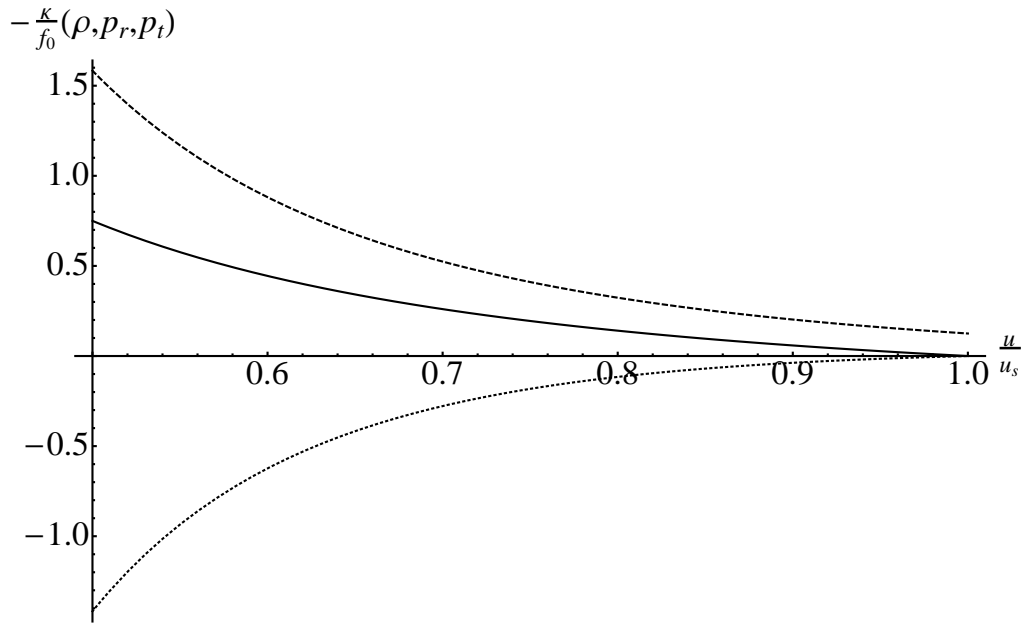}
\caption{The plot of effective density (thick line), radial pressure (dotted line), and tangential pressure (dashed line) in terms of $u$.}
\label{F1}
\end{figure}

The effective density and pressure profiles are plotted in figure (\ref{F1}). As it can be seen in the plot, the effective density and tangential pressure are positive and decreasing, while the radial pressure is negative and increasing in the halo. An interesting property of these profiles is that in terms of scaled distance $u/u_s$, the scaled density and pressure (scaled with $-f_0/\kappa$) are independent of any parameter and thus they are universal.  
 
\begin{figure}[h] 
\centering
\includegraphics[scale=0.5]{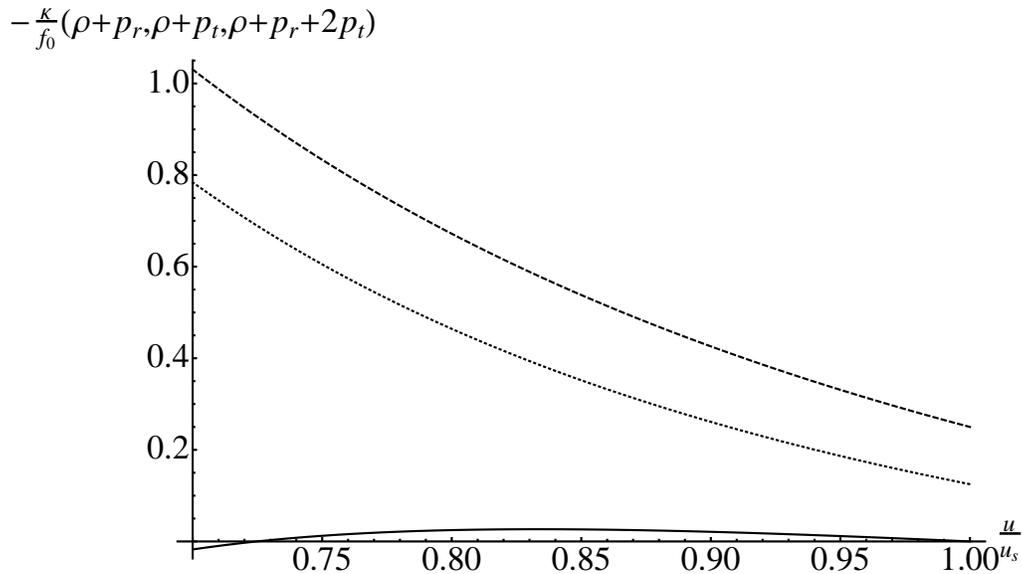}
\caption{The plot of $\rho+p_r$ (thick line), $\rho+p_t$ (dotted line), and $\rho+p_r+2p_t$ (dashed line) in terms of $u$.}
\label{F2}
\end{figure}

Let\rq{}s now  apply the energy conditions to this fluid. This provides us ranges of  $r$ in which energy conditions are satisfied. The result is plotted in figure (\ref{F2}) and summarized in Table (\ref{T1}). According to table (\ref{T1})  in the outer part of the halo, $\eta_0<\frac{u}{u_s}<1$ ($\eta_0\simeq 0.725$)\footnote{where $\eta_0$ is the value of $u/u_0$for which $\rho+p_r=0$. This leads to a cubic equation $4\eta_0^3-9\eta_0+5=0$ which has three real solutions $\eta_0=1$, $\eta_0\simeq -1.725$ and $\eta_0\simeq 0.725$. Only the last root is acceptable for our discussion.} the weak, null and strong energy conditions are satisfied but dominant energy condition is not satisfied anywhere. Thus the effective matter representing the halo is  not a normal kind of matter (named by M. Visser in \cite{matt}, because of the breakdown of DEC) and gravitation is attractive in region  $\eta_0<\frac{u}{u_s}<1$ (because of the validity of SEC). Again, as for figure (\ref{F1}), the domain of validity of energy conditions is also independent of the parameters of the model.
\begin{table}[h]
\begin{center}
\begin{tabular}{l|l}
Energy condition & Domain of validity\\
\hline
\hline
WEC ($\rho\ge 0;\ \ \rho+p_r>0;\ \ \rho+p_t>0$) & $\eta_0<\frac{u}{u_s}<1$\\
\hline
NEC ($\rho\ge 0;\ \ \rho+p_r\ge 0;\ \ \rho+p_t\ge 0$) & $\eta_0\le\frac{u}{u_s}\le 1$\\
\hline
SEC ($\rho+p_r\ge 0;\ \ \rho+p_t\ge 0;\ \ \rho+p_r+2p_t\ge 0$) & $\eta_0\le\frac{u}{u_s}\le 1$\\
\hline
DEC ($\rho\ge |p_r|;\ \ \rho\ge |p_t|$) &  Never\\
\end{tabular}
\end{center}
\caption{Energy conditions, $\eta_0\simeq 0.725$.}
\label{T1}
\end{table}

\begin{figure}[h]
\centering
\includegraphics[scale=0.5]{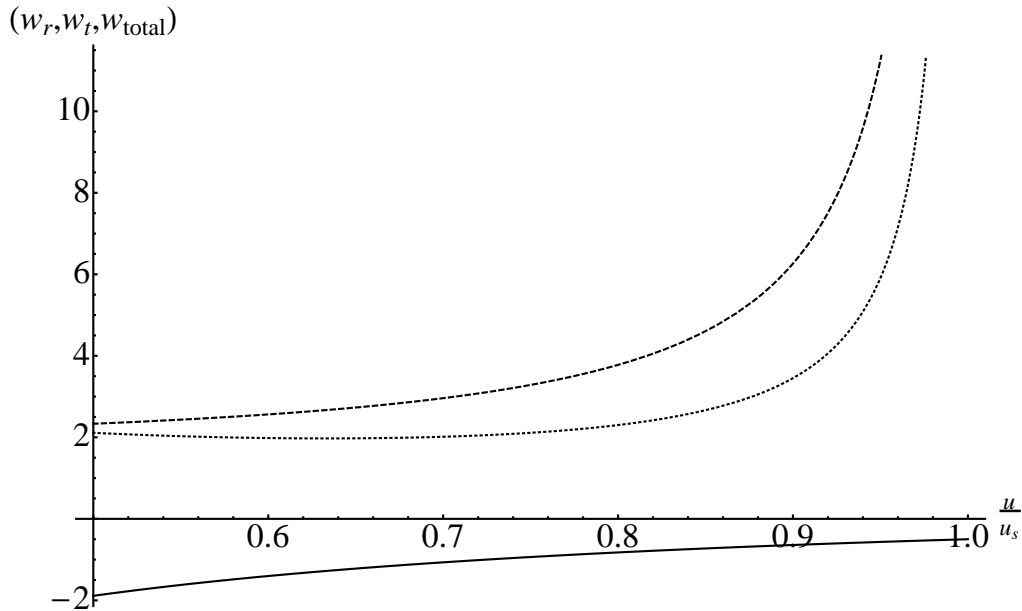}
\caption{The plot of $w_r$ (thick line), $w_t$ (dotted line), and $w_{\textrm{\tiny total}}$ (dashed line) in terms of $u$.}
\label{F3}
\end{figure}

Defining $\omega_r=p_r/\rho$, $\omega_t=p_t/\rho$ and $\omega_{total}=(p_r+2p_t)/\rho$, the plots of $\omega_r$, $\omega_t$ and $\omega_{total}$ are shown in figure (\ref{F3}). The equation of state parameters as functions of the scaled distance $u/u_s$ are universal and thus independent of the model's parameters. One can easily see that $\omega_{total}$ is of the order of magnitude $1$ for $\frac{u}{u_s}<0.9$.  This means that till near the surface of the halo the total pressure and density are of the same order. Near the surface of the halo the total pressure is very high. Thus in  this $f(R)$ model for dark matter, the halo obeys a non--Newtonian dynamics.

These results help us to compare $f(R)$ dark matter model with the scalar dark matter halo for which $p_t=-\rho$ \cite{mat}.  In that model there exists an arbitrary integration constant and the authors don't use the matching conditions to fix it. This is an important  difference between the model of \cite{mat} (and also \cite{rahman}) and ours. In \cite{mat} it is shown that this arbitrary constant should be smaller than $10^{-7}$ in order to have attractive gravity as well as being in the Newtonian regime.

Although the presented model can describe the flat rotation curves, the form of $f(R)$ should be written in a universal form, i.e. independent of the galaxies\rq{} parameters, to have a geometrical description of dark matter (This fact is adopted in a variety of similar works in the literature, e.g. \cite{X1}, while it is forgotten in some works \cite{sob,as}).  
  The relation (\ref{f}) can be written in a universal form in the domain of validity of our approximation. To see this let\rq{}s write the equations (\ref{44}) and (\ref{f}) using equations (\ref{const}) in the form:
\begin{equation}
\frac{R}{R_0}=\frac{1}{u^2}-\frac{2}{3u_s^2}
\label{z1}
\end{equation}
\begin{equation}
f(R)=R-R_1-\frac{5\sqrt{2}\alpha}{9\sqrt{3R_1}}(R+R_1)^{3/2}+\frac{\alpha R_1}{2}\left ( 1+\frac{3(R+R_1)}{2R_0} \right) \ln \left( \frac{R+R_1}{R_0}\right )
\label{z2}
\end{equation}
where we have set $C_3=1/2$ and $F_0=1$ for simplifying the relations, and $R_0=3\alpha/r_{\textrm{\tiny opt}}^2$ and $R_1=2\alpha/r_s^2$.
It should be noted that the radius of the halo, $r_s$, is very larger than the characteristic scale of the galaxy $r_{\textrm{\tiny opt}}$, i.e. $r_s>>r_{\textrm{\tiny opt}}$. Using the relation (\ref{z1}) one gets:
\begin{equation}
R\sim R_0>> R_1
 \end{equation}
  This leads to
\begin{equation}
f(R)\simeq R-\gamma R^{3/2}
\label{last1}
 \end{equation}
  where
  \begin{equation}
\gamma=\frac{5}{9\sqrt{3}}\sqrt{\alpha}r_s
 \end{equation}
Choosing $\gamma$ as a universal constant, we get a geometric description of flat rotation curves. The radius of halo is inversely proportional to square root of $\alpha$ or inversely proportional to $v_{tg}$.

According to (\ref{v}), the tangential velocity  is given by:
 \begin{equation}
 v_{tg}^2=\frac{243\gamma^2}{50r_s^2}\frac{1}{1+\frac{243\gamma^2}{25r_s^2}\ln \frac{r}{r_{\textrm{\tiny opt}}}}
  \end{equation}
  Therefore the tangential velocity depends on the universal constant $\gamma$ as well as on the parameters $r_s$ and $r_{\textrm{\tiny opt}}$ which are specific to each galaxy. 
\section{Lensing effect of halo}
In order to see how the halo affects the gravitational lensing, here we investigate the bending of a light ray passing through the space--time with the metric (\ref{met}), in which we are using the relations (\ref{ain}) and (\ref{x}).
 The geodesic equations for a light ray moving in the equatorial plane, $\theta=\pi/2$, can be written as:
\begin{equation}
A\frac{dt}{d\lambda}=k
\end{equation}
\begin{equation}
r^2\frac{d\phi}{d\lambda}=h
\end{equation}
\begin{equation}
0=\frac{k^2}{A}-\frac{X}{A}\left(\frac{dr}{d\lambda}\right)^2-\frac{h^2}{r^2}
\end{equation}
where $\lambda$ is an arbitrary affine parameter and $k$, $h$ are constants related to the energy and angular momentum of the photon. At the closest approach to origin, $r_0$, one has:
\begin{equation}
\frac{k^2}{h^2}=\frac{A(r_0)}{r_0^2}
\end{equation}
Introducing the new variable $s=1/r$, the orbit equation is then:
\begin{equation}
\left(\frac{ds}{d\phi}\right)^2=\frac{A(r_0)}{X(1/s)}\left(\frac{1}{r_0^2}-s^2\frac{A(1/s)}{A(r_0)}\right)
\end{equation}
Using the above equation we can evaluate the final angle at which the light ray propagates as: 
\begin{equation}
\Delta(\alpha)=2\int_0^{1/r_0} ds \sqrt{\frac{X(1/s)}{A(r_0)\left( \frac{1}{r_0^2}-s^2\frac{A(1/s)}{A(r_0)} \right)}}
\end{equation}
The deflection angle can be obtained by subtracting the no--gravity part from the above relation. For our case, no--gravity is given by $\alpha=0$. Therefore the deflection angle is given by: 
\begin{equation}
\delta=\Delta(\alpha)-\Delta(0)\simeq \alpha\left .\frac{d\Delta}{d\alpha}\right|_{\alpha=0}
\end{equation}
Let\rq{}s first assume that the closest point is inside the halo, then the deflection angle  is given by
\[
\Delta(\alpha)=2\int_0^{1/r_s} ds \sqrt{\frac{X^{\textrm{\tiny out}}(1/s)}{A^{\textrm{\tiny in}}(r_0)\left( \frac{1}{r_0^2}-s^2\frac{A^{\textrm{\tiny out}}(1/s)}{A^{\textrm{\tiny in}}(r_0)} \right)}}+
\]
\begin{equation}
 2\int_{1/r_s}^{1/r_0} ds \sqrt{\frac{X^{\textrm{\tiny in}}(1/s)}{A^{\textrm{\tiny in}}(r_0)\left( \frac{1}{r_0^2}-s^2\frac{A^{\textrm{\tiny in}}(1/s)}{A^{\textrm{\tiny in}}(r_0)} \right)}}
\end{equation}
in which $r_s$ is the radius of the surface of the halo. Using (\ref{aout}) and (\ref{xout})  a straightforward calculation yields:
\begin{equation}
\frac{\delta}{\alpha}=1+\left (\frac{u_0}{u_s} -\frac{u_s}{u_0}\right)\sqrt{1- \frac{u_0^2}{u_s^2}}
\label{last}
\end{equation}
 in which $u_0=\frac{r_0}{r_{\textrm{\tiny opt}}}$.
 
  On the other hand for the case where the closest approach is outside the halo, the propagation angle is given by:
 \begin{equation}
\Delta(\alpha)=2\int_0^{1/r_0} ds \sqrt{\frac{X^{\textrm{\tiny out}}(1/s)}{A^{\textrm{\tiny out}}(r_0)\left( \frac{1}{r_0^2}-s^2\frac{A^{\textrm{\tiny out}}(1/s)}{A^{\textrm{\tiny out}}(r_0)} \right)}}
 \end{equation}
 After doing a similar calculation, the result is just the familiar result of the Schwarzschield metric:
\begin{equation}
\frac{\delta}{\alpha}=\frac{u_s}{u_0}
\end{equation}

\begin{figure}[h]
\centering
\includegraphics[scale=0.5]{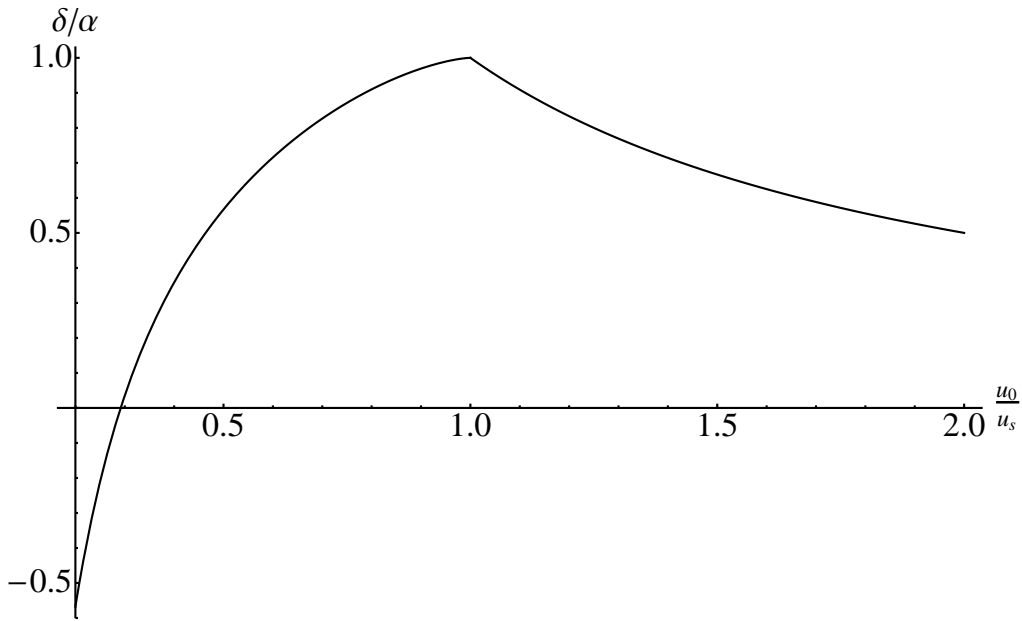}
\caption{The plot of light deflection angle in terms of $u_0/u_s$.}
\label{F4}
\end{figure}

The deflection angle is plotted in figure (\ref{F4}). An interesting feature is that for some values of $r_0$, the deflection angle can be negative, that is we have reflection instead of refraction.
For $u/u_s<\zeta_0$ the deflection angle is negative, where $\zeta_0\simeq 0.318$ is the only real root of equation (\ref{last}). It is interesting to note that the value of $r_0/r_s$ for which below it the deflection angle becomes negative is independent of the parameters of the model.

The appearance of negative deflection angle is not surprising. This is because of the fact that for $f(R)$ models, the energy conditions (on the effective density and pressure) can break. If one considers a congruence of geodesics in such models, there are cases in which this congruence is diverging instead of converging, see e.g. \cite{af}. In fact for our model, when one considers a sufficiently small $r_0$, $f(R)$ (given by equation (\ref{z2})) differs dramatically from $R$ and thus the strong energy condition breaks sufficiently and leads to a diverging congruence. The negative deflection angle can also be seen in some scalar--tensor models\cite{n1} and in higher dimensional theories\cite{n2}, in which again negative deflection angles are resulted from breakdown of effective strong energy condition leading to a negative tidal force. 

Here this effect is resulted from the negativity of the radial pressure. This does not mean that we have anti--gravity. 
In the region that the strong energy condition is satisfied the gravity is attractive. For the region of validity of strong energy condition the deflection angle has a peak and this means that a wide congruence of light rays are deflected to a narrow angle and hence the gravity is attractive. 
\section{Conclusions}
In this paper we have investigated the question of whether dark matter can have a geometrical nature. In the context of $f(R)$ theory of gravity, we found two solutions describing the dark halo of spiral galaxies. 
The first solution is exact but cannot be matched to the exterior Schwarzschild metric, while the second one is an approximate solution and can be matched to the exterior solution. 
The solution is completely determined using the asymptotic velocity of rotation curve, and the definition of halo surface as the surface of zero effective density and radial pressure.  
These are enough to fix the form of $f(R)$ and  the effective equation of state in the galaxy halo. 

Studying  the energy conditions in the halo, shows that all the energy conditions are satisfied for radial coordinate larger than a critical value except the dominant energy condition which is never satisfied. 
This critical value is $r\simeq 0.725 r_s$.
 
Generally whether the dark matter pressure is smaller than (or of the same order of)  its density or not, is an open question  and it depends on the nature of dark matter component\cite{fab}.
Since the total equation of state parameter in the presented model is larger than one throughout the halo, the effective dark matter obeys a non--Newtonian dynamics.

We also investigated the lensing effect of the halo in this model and saw that for some values of impact parameters the lensing effect can result in negative deflection angle, which means that we have reflection instead of refraction. This is because of the fact that the radial pressure is negative and in the inner regions it is so negative that it can exert such a force that leads to reflection.

There are some directions that one can generalize the results. First, although we did all the calculations up to first order in $\alpha$, it is not very difficult to go to any order of perturbation. Second, we have used the asymptotic form of rotation curves, i.e. the flat portion, but in principle the whole rotation curve can be adopted for the purpose of obtaining an $f(R)$ model. The calculations would be very cumbersome and thus using the weak field approximation helps. Finally, it should be noted that although we used the Schwarzschild solution for the exterior region, another choice may be some appropriate solution of $f(R)$ models used in describing dark energy for the outer space--time. In this way, it might be possible to find an $f(R)$ model appropriate both for galactic and cosmic scales. The question of possibility of finding such a model needs detailed calculations.

Therefore, the presented model should not be viewed as a universal model. It is only valid for describing the galactic halos. In fact using the obtained form of $f(R)$ in equation (\ref{last1}) one can see that $df/dR$ would be negative below $r/r_s\sim 5\alpha/6$. Setting $r_s=10 \textrm{ kpc}$ e.g., below $r\sim 100 \textrm{ light hours}$ $df/dR$ is negative. Also the second derivative $d^2f/dR^2=-(3/4)\gamma R^{-1/2}$ is always negative. Therefore we have problems with stability and thus the model should not be extrapolated to other scales.  

\textbf{Acknowledgment} This work is  supported by a grant from university of Tehran.

\end{document}